\documentstyle[12pt]{article}

\begin{document}

\baselineskip=24pt 

\begin{titlepage}

\begin{flushright}
UR--1446 \\
November 1995
\end{flushright}

\vspace{0.24in}

\begin{center}

{\large\bf A Two Higgs Doublet Model for the Top Quark}

\vspace{0.36in}

Ashok Das
and
Chung Kao\footnote{Internet Address: Kao@Urhep.pas.rochester.edu}

{\sl Department of Physics and Astronomy, University of Rochester, \\
Rochester, NY 14627, USA}

\end{center}

\vspace{0.36in}

\begin{abstract}

A two Higgs doublet model
with special Yukawa interactions for the top quark and
a softly broken discrete symmetry in the Higgs potential is proposed.
In this model, the top quark is much heavier than the other quarks
and leptons because it couples to a Higgs doublet
with a much larger vacuum expectation value.
The electric dipole moment (EDM) of the electron is evaluated
with loop diagrams of the third generation fermions
as well as the charm quark.
The electron EDM is significantly enhanced
for a naturally large $\tan\beta \equiv |v_2|/|v_1|$.

\end{abstract}

\end{titlepage}

\newpage

\section{Introduction}

Recently, the top quark with a large mass has been observed
at the Fermilab Tevatron \cite{CDF,Dzero}.
Since the top quark is much heavier than all other known fermions,
it might provide some clue to unravel
the mystery of electroweak symmetry breaking.
In the Standard Model (SM) of electroweak interactions,
only one Higgs doublet is required to generate masses for fermions
as well as gauge bosons.
A neutral CP-even Higgs boson ($H^0$)
remains after spontaneous symmetry breaking.
The mass of a quark or a lepton is given by its Yukawa coupling constant
with the Higgs boson times the vacuum expectation value (VEV)
of the Higgs field.

A two Higgs doublet model \cite{Georgi} has doublets $\phi_1$
and $\phi_2$ with VEVs $v_1/\sqrt{2}$ and $v_2/\sqrt{2}$.
There remain five `Higgs bosons' after symmetry breaking:
a pair of singly charged Higgs bosons $H^{\pm}$, two neutral CP-even scalars
$H_1$ and $H_2$, and a neutral CP-odd pseudoscalar $A$.
Several two Higgs doublet models have been suggested
with different Yukawa interactions for fermions and spin-0 bosons.
In Model I \cite{Model1}, the different mass scales of the fermions and
the gauge bosons are set by different Higgs VEVs.
In Model II \cite{Model2a,Model2b},
one Higgs doublet couples to down-type quarks and charged leptons
while another doublet couples to up-type quarks and neutrinos.
A more recent model \cite{Ruiming91} was proposed to explain
why $m_u/m_d$ is so much smaller than $m_c/m_s$ and $m_t/m_b$.

We propose that the top quark is much heavier than the other quarks
and leptons, because,
in the three known fermion generations,
it is the only elementary fermion getting a mass
from a much larger VEV of a second Higgs doublet.
This model has a few interesting features:
(1) The top quark is naturally heavier than other quarks and leptons.
(2) The ratio of the Higgs VEVs, $\tan\beta \equiv |v_2|/|v_1|$,
is naturally large in this model, which highly enhances the Yukawa couplings
of the lighter quarks and leptons with the Higgs bosons.
(3) There are flavor changing neutral Higgs (FCNH) interactions.

A significant electric dipole moment (EDM) for the electron and the neutron
can be generated if Higgs boson exchange generates CP violation
\cite{Steve89}-\cite{Barr&Zee}.
In this paper, we discuss the effect of a large $\tan\beta$
with CP violation from the Higgs exchange.
The phenomenology of FCNH interactions as well as Yukawa interactions
for the charged Higgs boson will be presented in the near future.

In Section II, we present the Lagrangian density for the  Yukawa interactions
of this model.
As a first application to phenomenology,
the electron EDM is evaluated in Section III,
with contributions from fermion loops including $t$, $b$, $\tau$, and $c$,
and CP violation generated from neutral Higgs exchange.
Promising conclusions are drawn in Section IV.

\section{Yukawa Interactions}

We choose the Lagrangian density of Yukawa interactions
to be of the following form
\begin{eqnarray}
{\cal L}_Y
& = & -\sum_{m,n=1}^{3} \bar{L}^m_L \phi_1 E_{mn} l^n_R
      -\sum_{m,n=1}^{3} \bar{Q}^m_L \phi_1 F_{mn} d^n_R
       \nonumber \\
&   & -\sum_{\alpha=1}^{2} \sum_{m=1}^{3} \bar{Q}^m_L
                           \tilde{\phi_1} G_{m\alpha} u^\alpha_R
      -\sum_{m=1}^{3} \bar{Q}^m_L \tilde{\phi_2} G_{m3} u^3_R
       \nonumber \\
&   & +{\rm H.c.}
\end{eqnarray}
where
\begin{eqnarray}
\phi_\alpha
& = & \left( \begin{array}{c} \phi_\alpha^+ \\
                              \frac{ v_\alpha +\phi_\alpha^0 }{ \sqrt{2} }
             \end{array} \right) \nonumber \\
\tilde{\phi}_\alpha
& = &
\left( \begin{array}{c} \frac{ v_\alpha^* +{\phi_\alpha^0}^* }{ \sqrt{2} } \\
                             -\phi_\alpha^-
\end{array} \right), \,\, \phi_\alpha^- = {\phi_\alpha^+}^*, \,\, \alpha = 1,2
\end{eqnarray}
and
\begin{eqnarray}
L^m_L & = & \left( \begin{array}{c} \nu_l \\ l \end{array} \right)^m_L,
            \nonumber \\
Q^m_L & = & \left( \begin{array}{c} u \\ d \end{array} \right)^m_L,
            \,\, m= 1,2,3
\end{eqnarray}
$l^m$, $d^m$, and $u^m$ are the leptons, the down-type quarks
and the up-type quarks in the gauge eigenstates.
This Lagrangian respects a discrete symmetry,
\begin{eqnarray}
\phi_1 &\to& -\phi_1, \,\, \phi_2 \to +\phi_2, \nonumber \\
l^m_R  &\to& -l^m_R,  \,\, d^m_R  \to -d^m_R, \,\, u^\alpha_R  \to -u^\alpha_R,
             \nonumber \\
L^m_L  &\to& +L^m_L,  \,\, Q^m_L  \to +Q^m_L, \,\, u^3_R  \to +u^3_R,
\end{eqnarray}
with $m = 1,2,3$ and $\alpha = 1,2$.
In this model,
only the top quark has bilinear couplings to the doublet $\phi_2$,
while all other quarks and leptons have bilinear couplings
to the doublet $\phi_1$.
The Yukawa interactions of the down-type quarks and leptons
with neutral Higgs bosons are the same as those in Model II.

The fermion masses are generated when the $\phi$'s have developed
VEVs, $<\phi_1> = v_1/\sqrt{2}$ and $<\phi_2> = v_2/\sqrt{2}$,
which can both be complex.
We propose that $|v_2| \gg |v_1|$
and $\tan\beta \equiv |v_2|/|v_1|$ is close to $m_t/m_b$,
so that $m_t$ is much larger than $m_b$.

The mass terms for the up quarks are
\begin{equation}
{\cal L}^U_M = -\sum_{m,n=1}^{3} \bar{u}^m_L \tilde{M}_U u^n_R,
\end{equation}
where
\begin{eqnarray}
\tilde{M}_U = \frac{v_1^*}{\sqrt{2}}
\left( \begin{array}{ccc}
G_{11} & G_{12} & (v_2^*/v_1^*) G_{13} \\
G_{21} & G_{22} & (v_2^*/v_1^*) G_{23} \\
G_{31} & G_{32} & (v_2^*/v_1^*) G_{33}
\end{array} \right)
\end{eqnarray}
and $u^m$, $m = 1,2,3$ are the gauge eigenstates.
Let us introduce unitary transformations
\begin{eqnarray}
\left( \begin{array}{c}
u^1 \\ u^2 \\ u^3
\end{array} \right)_{L,R}
= U_{L,R}
\left( \begin{array}{c}
u \\ c \\ t
\end{array} \right)_{L,R}
\end{eqnarray}
such that
\begin{eqnarray}
U_L^\dagger \tilde{M}_U U_R =
U_R^\dagger \tilde{M}_U^\dagger U_L =
\left( \begin{array}{ccc}
m_u & 0 & 0 \\
0 & m_c & 0 \\
0 & 0 & m_t
\end{array} \right)
\end{eqnarray}
where $u,c$, and $t$ are the mass eigenstates.

The neutral Yukawa interactions of the up quarks are
\begin{eqnarray}
{\cal L}^U_N
& = &-\sum_{u = u,c,t}
      m_u  \bar{u}_L u_R \frac{ {\phi_1^0}^* }{ v_1^* }
     -\sum_{ab} \bar{u}^a_L {\Sigma}_{ab} u^b_R
      \left( \frac{ {\phi_2^0}^* }{ v_2^* }
            -\frac{ {\phi_1^0}^* }{ v_1^* } \right), \nonumber \\
&   & +{\rm H.c.}
\end{eqnarray}
where $u^{a,b} = u,c,t$ and
\begin{eqnarray}
\Sigma
& = &
\left( \begin{array}{ccc}
m_u & 0   & 0 \\
0   & m_c & 0 \\
0   & 0   & m_t
\end{array} \right)
U_R^\dagger
\left( \begin{array}{ccc}
0 & 0 & 0 \\
0 & 0 & 0 \\
0 & 0 & 1
\end{array} \right)
U_R \nonumber \\
& = &
\left( \begin{array}{ccc}
m_u |z_{31}|^2      & m_u z_{31}^* z_{32} & m_u z_{31}^* z_{33} \\
m_c z_{32}^* z_{31} & m_c |z_{32}|^2      & m_c z_{32}^* z_{33} \\
m_t z_{33}^* z_{31} & m_t z_{33}^* z_{32} & m_t |z_{33}|^2
\end{array} \right)
\end{eqnarray}
where
\begin{equation}
U_R =
\left( \begin{array}{ccc}
 z_{11} & z_{12} & z_{13} \\
 z_{21} & z_{22} & z_{23} \\
 z_{31} & z_{32} & z_{33}
\end{array} \right)
\end{equation}
To a good approximation, the unitary matrix $U_R$
has the following form
\begin{equation}
U_R =
\left( \begin{array}{rrr}
 \cos\phi   & -\sin\phi   & -\cos\phi\epsilon_1^* +\sin\phi\epsilon_2^* \\
 \sin\phi   &  \cos\phi   & -\sin\phi\epsilon_1^* -\cos\phi\epsilon_2^* \\
 \epsilon_1 &  \epsilon_2 & 1
\end{array} \right)
\end{equation}
We have introduced two small parameters\footnote{This is similar to
what was also suggested for the $U_L$ in a recent model \cite{Hill}
with Topcolor dynamics.}
$\epsilon_1 = |\epsilon_1|e^{i\delta_1}$ and
$\epsilon_2 = |\epsilon_2|e^{i\delta_2}$, with
$|\epsilon_1| \sim  m_u/m_t$ and $|\epsilon_2| \sim  m_c/m_t$.
We will keep terms only to the first order in the $\epsilon$'s in our analysis.

Introducing a transformation, which takes the two Higgs doublets
to the gauge eigenstates ($\Phi_1$ and $\Phi_2$),
such that $<\Phi_1> = v/\sqrt{2}$, $<\Phi_2> = 0$, we have
\begin{eqnarray}
\phi_1 & = & ( \cos \beta \Phi_1 -\sin \beta \Phi_2 ), \nonumber \\
\phi_2 & = & ( \sin \beta \Phi_1 +\cos \beta \Phi_2 ) e^{i\theta} , \\
\Phi_1  & = &
\left( \begin{array}{c}
G^+ \\ \frac{v +H_1 +iG^0}{\sqrt{2}}
\end{array} \right), \nonumber \\
\Phi_2  & = &
\left( \begin{array}{c}
H^+ \\ \frac{ H_2 +iA}{\sqrt{2}}
\end{array} \right), \\
v & = & \sqrt{ |v_1|^2 +|v_2|^2},
\end{eqnarray}
where $G^\pm$ and $G^0$ are Goldstone bosons,
$H^\pm$ are singly charged Higgs bosons,
$H_1$ and $H_2$ are CP-even scalars,
and $A$ is a CP-odd pseudoscalar.
Without loss of generality, we will take
$<\phi_1> = v_1/\sqrt{2}$ and $<\phi_2> = v_2 e^{i\theta}/\sqrt{2}$,
with $v_1$ and $v_2$ real and $\tan\beta \equiv v_2/v_1$.

The neutral Yukawa interactions of the quarks now become
\begin{eqnarray}
{\cal L}^N_Y
& = &-\sum_{d=d,s,b} m_d \bar{d}d
     -\sum_{u=u,c,t} m_u \bar{u}u
      \nonumber \\
&   &-\sum_{d=d,s,b}\frac{m_d}{v}\bar{d}d ( H_1 -\tan\beta H_2 )
     -i\sum_{d=d,s,b}\frac{m_d}{v}\bar{d}\gamma_5 d  ( G^0 -\tan\beta A )
      \nonumber \\
&   &-\frac{m_u}{v}\bar{u}u [ H_1 -\tan\beta H_2 ]
     -\frac{m_c}{v}\bar{c}c [ H_1 -\tan\beta H_2 ] \nonumber \\
&   &-\frac{m_t}{v}\bar{t}t [ H_1 +\cot\beta H_2 ] \nonumber \\
&   &+i\frac{m_u}{v}\bar{u} \gamma_5 u [ G^0 -\tan\beta A ]
     +i\frac{m_c}{v}\bar{c} \gamma_5 c [ G^0 -\tan\beta A ] \nonumber \\
&   &+i\frac{m_t}{v}\bar{t} \gamma_5 t [ G^0 +\cot\beta A ] \nonumber \\
&   &+{\cal L}_{\rm FCNH}
\end{eqnarray}
where ${\cal L}_{\rm FCNH}$ are the terms that will generate
flavor changing neutral Higgs interactions,
\begin{eqnarray}
{\cal L}_{\rm FCNH}
& = & \{ \epsilon_1^* \bar{u}[ -(m_u+m_t) +(m_t-m_u)\gamma_5 ] t \nonumber \\
&   &   +\epsilon_1   \bar{t}[ -(m_u+m_t) -(m_t-m_u)\gamma_5 ] u \nonumber \\
&   &   +\epsilon_2^* \bar{c}[ -(m_c+m_t) +(m_t-m_c)\gamma_5 ] t \nonumber \\
&   &   +\epsilon_2   \bar{t}[ -(m_c+m_t) -(m_t-m_c)\gamma_5 ] c \}
      \times (\frac{H_2}{v\sin 2\beta})
      \nonumber \\
&   &+i\{ \epsilon_1^* \bar{u}[-(m_t-m_u) +(m_u+m_t)\gamma_5 ] t \nonumber \\
&   &    +\epsilon_1   \bar{t}[+(m_t-m_u) +(m_u+m_t)\gamma_5 ] u \nonumber \\
&   &    +\epsilon_2^* \bar{c}[-(m_t-m_c) +(m_c+m_t)\gamma_5 ] t \nonumber \\
&   &    +\epsilon_2   \bar{t}[+(m_t-m_c) +(m_c+m_t)\gamma_5 ] c \}
      \times (\frac{A}{v\sin 2\beta}).
\end{eqnarray}
There is no FCNH interactions between the up and the charm quarks.

\section{The Electron Electric Dipole Moment}

The experimental bound on the EDM of the electron is
$d_e = (-2.7\pm 8.3) \times 10^{-27}$ e$\cdot$cm \cite{EDM}.
The electron EDM ($d_e$) in the SM, generated from
the Cabibbo-Kobayashi-Maskawa (CKM) phase,
has been found to be extremely small \cite{Hoogeveen}.
It is too small to be observed.

In a multi-Higgs-doublet model, for flavor symmetry to be conserved naturally
to a good degree, a discrete symmetry \cite{Glashow} is usually required.
In a model with two Higgs doublets only, there is no
CP violation from the Higgs sector if the discrete symmetry enforcing
the natural flavor conservation were exact. By letting this symmetry
be broken by soft terms, CP violation can be introduced while the
flavor changing interaction can still be kept at an acceptably low level
\cite{Branco,Jiang}.

In two Higgs doublet models, a significant electron EDM
can be generated if Higgs boson exchange mediates CP violation.
There are contributions from two-loop diagrams
with the top quark \cite{Barr&Zee},
the gauge bosons \cite{Barr&Zee}-\cite{Darwin}
and the charged Higgs boson \cite{Kao&Xu}.
The contributions from the $b$ quark
and the $\tau$ dominate for $\tan\beta$ larger than about 10
with the same Yukawa interactions as those of the Model II.
In our model, in addition,
even the $c$ quark loop produces a large electron EDM
for a large $\tan\beta$.

Adopting Weinberg's parameterization \cite{Steve90}
and applying the identity $v^2 = (\sqrt{2} G_F)^{-1}$,
we can write the following neutral Higgs exchange propagators as
\begin{eqnarray}
< H_1 A >_q
& = &
    \frac{\sin 2\beta}{2} \sum_n \frac{ {\rm Im}Z_{0n} }{ q^2-m_n^2 }
    \nonumber \\
& = &
    \frac{1}{2} \sum_n
    \frac{-\cos^2 \beta \cot \beta{\rm Im}\tilde{Z}_{1n}
          +\sin^2 \beta \tan \beta{\rm Im}\tilde{Z}_{2n} }{ q^2-m_n^2 }
    \nonumber \\
< H_2 A >_q
& = &
    \frac{1}{2} \sum_n
    \frac{ \cos 2\beta {\rm Im}Z_{0n} -{\rm Im}\tilde{Z}_{0n} }{ q^2-m_n^2 }
    \nonumber \\
& = &
    \frac{1}{2} \sum_n
    \frac{ \cos^2 \beta {\rm Im}\tilde{Z}_{1n}
          +\sin^2 \beta {\rm Im}\tilde{Z}_{2n} }{ q^2-m_n^2 },
\label{eq:H&A}
\end{eqnarray}
where the summation is over all the mass eigenstates of neutral Higgs
bosons. We will approximate the above expressions by assuming that
the sums are dominated by a single neutral Higgs boson of mass $m_0$,
and drop the sums and indices $n$ in Eq.~\ref{eq:H&A} hereafter.

There are interesting relations among the CP violation parameters,
\begin{eqnarray}
{\rm Im} Z_0 +{\rm Im} \tilde{Z}_0 & = & -\cot^2 \beta {\rm Im} \tilde{Z}_1,
                                          \nonumber \\
{\rm Im} Z_0 -{\rm Im} \tilde{Z}_0 & = & +\tan^2 \beta {\rm Im} \tilde{Z}_2.
\end{eqnarray}
Weinberg has shown \cite{Steve90} that
\begin{eqnarray}
|{\rm Im}Z_1| \le (1/2) |\tan\beta| ( 1+\tan^2\beta)^{1/2}, \nonumber \\
|{\rm Im}Z_2| \le (1/2) |\cot\beta| ( 1+\cot^2\beta)^{1/2},
\end{eqnarray}
with unitarity constraints.

The Feynman diagrams of fermion loops contributing to the electron EDM
are shown in Figure 1.
The diagrams with the intermediate $Z$ boson are highly suppressed
by the vector part of the $Ze^+e^-$ couplings.
Therefore, we consider only the diagrams involving an intermediate $\gamma$.
In our analysis, we will take $m_t = 175$ GeV, $m_b = 4.8$ GeV,
$m_\tau = 1.777$ GeV, $m_c = 1.4$ GeV,
and the fine structure constant $\alpha = 1/137$.

The top-loop contribution is \cite{Barr&Zee}\footnote{
In Eq.~(2) of this reference, ${\rm Im} Z_1$ should be $-{\rm Im} Z_1$.}
\begin{eqnarray}
\left( \frac{d_e}{e} \right)^{ t-{\rm loop} }
& = & -\frac{16}{3}
    \frac{ m_e \alpha \sqrt{2} G_F }{ (4\pi)^3 }
    \{ [ f(\rho_t) +g(\rho_t) ] {\rm Im}Z_0
      +[ g(\rho_t) -f(\rho_t) ] {\rm Im}\tilde{Z}_0 \}, \\
& = & -\frac{16}{3}
       \frac{ m_e \alpha \sqrt{2} G_F }{ (4\pi)^3 }
       [ -g(\rho_t) \cot^2 \beta {\rm Im}\tilde{Z}_1
         +f(\rho_t) \tan^2 \beta {\rm Im}\tilde{Z}_2 ],
\label{eq:top}
\end{eqnarray}
where $\rho_t = m_t^2/m_0^2$ and the functions $f$ and $g$ are defined as
\begin{eqnarray}
f(r) & \equiv & \frac{r}{2} \int_0^1 dx \frac{ 1-2x(1-x) }{ x(1-x)-r }
                \ln\left[ \frac{ x(1-x) }{r} \right], \nonumber \\
g(r) & \equiv & \frac{r}{2} \int_0^1 dx \frac{1}{ x(1-x)-r }
                \ln\left[ \frac{ x(1-x) }{r} \right].
\label{eq:f&g}
\end{eqnarray}
For $m_t \sim m_0$,
\begin{equation}
(d_e)^{ t-{\rm loop} }\approx -6.5 \times 10^{-27}\left[ {\rm Im} Z_0
+0.17 {\rm Im}\tilde{Z}_0\right] {\rm e}\cdot{\rm cm}.
\end{equation}
In Ref.~\cite{Ruiming90},  the fine structure constant $\alpha$ was taken
to be 1/128, therefore, our numerical data for the $t$ and $W$ loops are
(128/137) times smaller.

In our model, the EDM generated from the $b$ and the $\tau$ loops is
\begin{eqnarray}
\left( \frac{d_e}{e} \right)^{ b,\tau-{\rm loop} }
& = &-( 4N_c Q^2 \tan^2\beta )
      \frac{ m_e \alpha \sqrt{2} G_F }{ (4\pi)^3 }
      [ f(\rho_f) +g(\rho_f) ] ( {\rm Im}Z_0 +{\rm Im}\tilde{Z}_0 ), \\
& = &+( 4N_c Q^2 )
      \frac{ m_e \alpha \sqrt{2} G_F }{ (4\pi)^3 }
      \{ [ f(\rho_f) +g(\rho_f) ] {\rm Im} \tilde{Z}_1 \},
\label{eq:bottom}
\end{eqnarray}
The $N_c$ is the color factor and $Q$ is the charge.
For the $b$ and the $\tau$, $4N_c Q^2$ is equal to
$4/3$ and $4$ respectively.
The electron EDM generated from the $c$ loop is
\begin{eqnarray}
\left( \frac{d_e}{e} \right)^{ c-{\rm loop} }
& = & -( \frac{16}{3} \tan^2\beta )
       \frac{ m_e \alpha \sqrt{2} G_F }{ (4\pi)^3 }
       [ f(\rho_c) -g(\rho_c) ] ( {\rm Im}Z_0 +{\rm Im}\tilde{Z}_0 ), \\
& = & +( \frac{16}{3} )
      \frac{ m_e \alpha \sqrt{2} G_F }{ (4\pi)^3 }
      \{ [ f(\rho_c) -g(\rho_c) ] {\rm Im} \tilde{Z}_1 \},
\label{eq:charm}
\end{eqnarray}
where $\rho_f = m_f^2/m_0^2$ and $\rho_c = m_c^2/m_0^2$.
The difference between the contributions from the $c$-loop
and the $b$-loop comes from a relative sign between the $Ab\bar{b}$
and the $Ac\bar{c}$ couplings.

In Figures 2, we present
the electron EDM from heavy fermion loops (${d_e}^{ f-{\rm loop} }$),
in units of (a) Im $Z_0$ and (b) Im $\tilde{Z}_0$,
as a function of $m_0$, with $\tan\beta =$ 20,
where $m_0$ is the mass of the lightest physical spin-0 boson.
It is clear that for $m_0 < 200$ GeV and $\tan\beta >$ 10,
the fermion loops of the $b$ and the $\tau$ become dominant.

In Figures 3, we present
the electron EDM from heavy fermion loops (${d_e}^{ f-{\rm loop} }$),
in units of (a) Im $Z_0$ and (b) Im $\tilde{Z}_0$,
as a function of $\tan\beta$ with $m_0 = M_W = 80$ GeV.
For a large $\tan\beta$, the electron EDM
from the $W$-loop\cite{Ruiming90} is
\begin{equation}
d_e^{ W-{\rm loop} }
= +2.1 \times 10^{-26} {\rm Im} Z_0 \,\, {\rm e}\cdot {\rm cm}
\end{equation}
For $\tan\beta >$ 20 and $m_0 \sim M_W$, the fermion loops of the $b$,
and the $\tau$ become dominant.

There are several interesting aspects to note from the different contributions.
(1) The contributions from the $b$, the $\tau$ and the $c$ loops
are proportional to (Im $Z_0$ +Im $\tilde{Z}_0$).
(2) For the the $t$-loop, the coefficient of the Im $\tilde{Z}_0$
is much smaller than that of the Im $Z_0$.
(3) The $W$-loop does not contribute to the Im $\tilde{Z}_0$ term.
(4) The charm loop has the same sign as that of the $W$ and
the charged Higgs boson loops.
(5) For a large $\tan\beta$, the $c$-loop contribution can be larger than
that of the $t$-loop.

\section{Conclusions}

The model for Yukawa interactions proposed in this paper
has several interesting features:
(1) The top quark is naturally heavier than other quarks and leptons.
(2) The ratio of the Higgs VEVs, $\tan\beta \equiv v_2/v_1$,
is naturally large, which highly enhances the Yukawa couplings
of the bottom quark ($b$), the tau lepton ($\tau$),
and even the charm quark ($c$), with the Higgs bosons.
(3) There are flavor changing neutral Higgs (FCNH) interactions.

The electron EDM from loop diagrams of the bottom quark ($b$),
the tau lepton ($\tau$), and even the charm quark ($c$),
can be significantly enhanced with a large $\tan\beta$.
More precise experiments for the electron EDM will set bounds on
the $\tan\beta$ and the CP violation parameters,
the ${\rm Im}Z_0$, ${\rm Im}\tilde{Z}_0$, ${\rm Im}\tilde{Z}_1$, and
${\rm Im}\tilde{Z}_2$.
We might be able to unravel the mystery of electroweak symmetry breaking
and CP violation with the same `stone'.

\section*{Acknowledgments}

This research was supported in part
by the U.~S. Department of Energy grant DE-FG02-91ER40685.

\newpage
%

\newpage
\section*{Figures}

\begin{enumerate}

\item
Feynman diagrams for fermion loops contributing to
the electric dipole moment of the electron.
\label{fig:EDM}
%

\item
The electron EDM from fermion loops ${d_e}^{ f-{\rm loop} }$
in units of (a) Im $Z_0$ and (b) Im $\tilde{Z}_0$,
as a function of $m_0$, with $\tan\beta = 20$, for
the $t$-loop (dash),
the $b$-loop (dash-dot),
the $\tau$-loop (dot),
and the $c$-loop (dash-dot-dot),
where $m_0$ is the mass of the lightest physical spin-0 boson.
\label{fig:MH0}
%

\item The electron EDM from fermion loops ${d_e}^{ f-{\rm loop} }$
in units of (a) Im $Z_0$ and (b) Im $\tilde{Z}_0$,
as a function of $\tan\beta$ with $m_0 = 80$ GeV, for
the $t$-loop (dash),
the $b$-loop (dash-dot),
the $\tau$-loop (dot),
and the $c$-loop (dash-dot-dot).
\label{fig:tanb}

\end{enumerate}

\end{document}